\documentclass[11pt,a4paper,colorlinks=true,urlcolor=blue,anchorcolor=blue,citecolor=blue,filecolor=blue,linkcolor=blue,linktocpage=true,pdfproducer=medialab,pdfa=true]{article}
\usepackage{jcappub}
\usepackage{xfrac,float,cancel,booktabs}

\usepackage[normalem]{ulem}
\usepackage{soul}
\setstcolor{red}

\usepackage{graphicx}
\usepackage{dcolumn}
\usepackage{bm}
\usepackage{braket}
\usepackage{color}
\usepackage{orcidlink}

\title{The Lifespan of our Universe}

\author[a,1]{Hoang Nhan Luu\,\orcidlink{0000-0001-9483-1099}\note{Present address: Department of Physics \& Astronomy, University of New Hampshire Durham, NH 03824, USA.}}
\author[b,2]{Yu-Cheng Qiu\,\orcidlink{0000-0002-9008-4564}\note{Present address: Department of Physics, City University of Hong Kong, Kowloon, Hong Kong S.A.R. China.}}
\author[c,d]{and S.-H. Henry Tye\,\orcidlink{0000-0002-4386-0102}}

\affiliation[a]{Donostia International Physics Center, Basque Country UPV/EHU, San Sebastian, E-48080, Spain}
\affiliation[b]{Tsung-Dao Lee Institute, Shanghai Jiao Tong University, Shanghai, 201210, China}
\affiliation[c]{Department of Physics, Cornell University, Ithaca, NY 14853, USA}
\affiliation[d]{Department of Physics, The Hong Kong University of Science and Technology, Hong Kong S.A.R. China}

\emailAdd{hoang.luu@dipc.org}
\emailAdd{ethanqiu@sjtu.edu.cn}
\emailAdd{tye.henry@gmail.com}

\abstract{

The Dark Energy Survey (DES) and the Dark Energy Spectroscopic Instrument (DESI) measurements claim that the dark energy equation of state $w \ne -1$. This observation can be explained by the axion Dark Energy (aDE) model of an ultralight axion plus a cosmological constant $\Lambda$. Despite a relatively large degeneracy, there is a high probability that $\Lambda <0$. This negative $\Lambda$ leads the universe to end in a big crunch. Using the best-fit values of the model as a benchmark, we find the lifespan of our universe to be 33 billion years.

}

\arxivnumber{2506.24011}


\begin{document}

\maketitle

\section{Introduction}

The age of our universe is of foundational importance in cosmology. Since the establishment of the Big Bang Theory, we know it is finite. In the $\Lambda$CDM model, together with the Cosmic microwave background (CMB) and Baryonic Acoustic Oscillation (BAO) and Supernova (SN) data, its age is now determined to be $13.8$ billion years~\cite{Planck:2018vyg}. The next question is about its future; more specifically, its lifespan. In this paper, we address this issue.

Among various speculations, there are two simplest and so most likely scenarios where the fate of the universe is dictated by the value of the cosmological constant $\Lambda$ (positive, zero or negative $\Lambda$ corresponds respectively to de Sitter (dS), Minkowski, or anti-de Sitter (AdS) space): 

\begin{itemize}
    \item The universe continues its present expansion forever if $\Lambda \ge 0$, i.e., with an infinite lifespan.
    \item It eventually reaches a big crunch if $\Lambda < 0$, when the cosmic scale factor $a$ of the universe collapses to zero. 
\end{itemize} 

If the observed dark energy is due to $\Lambda$ alone, earlier cosmological observations imply that $\Lambda >0$, so the universe continues to expand forever, with the dark energy equation of state (EoS) $w=-1$. 
However, the recent Dark Energy Survey (DES) and the Dark Energy Spectroscopic Instrument (DESI) measurements~\cite{DES:2025bxy,DESI:2025zgx} claim that, at 4.2 $\sigma$ level, the dark energy is dynamical; that is, the dark energy EoS $w\ne -1$. It is known that a simple ultra-light axion may explain this behavior \cite{Frieman:1995pm,Waga:2000ay,Ng:2000di,Kawasaki:2001bq,Caldwell:2005tm,Hall:2005xb,Arvanitaki:2009fg,Hlozek:2014lca,Smer-Barreto:2015pla,Choi:2021aze,Bhattacharya:2024kxp,Tada:2024znt,Yin:2024hba,Wolf:2024eph,Abreu:2025zng,Shajib:2025tpd,Urena-Lopez:2025rad,Lin:2025gne,deSouza:2025rhv,Cline:2025sbt}. 
In addition to such an axion, the axion-dark energy (aDE) model~\cite{Luu:2025fgw} introduces a cosmological constant $\Lambda$~\footnote{See also refs.~\cite{Cardenas:2002np,Dutta:2018vmq,Visinelli:2019qqu,Ruchika:2020avj,Sen:2021wld,Murai:2025msx,Nakagawa:2025ejs}. Other implications on the $\Lambda<0$ case have been studied in refs.~\cite{Calderon:2020hoc,Adil:2023ara,Menci:2024rbq}.}, which happens to be a crucial feature. In applying this aDE model to the 
DES measurements, we find that the universe prefers to have a negative cosmological constant, i.e.,  $\Lambda < 0$. In this case, 
 $a$ reaches a maximum and then collapses to zero, ending in a big crunch. This big crunch defines the end of the universe. 

In comparing to the 
BAO + SN data from DES and other constraints from CMB, BBN, age of the universe, 
ref.\cite{Luu:2025fgw} finds that there is a rather large degeneracy in the parameter space of the aDE model, where $\Lambda <0$ is most likely, implying that the universe has a finite lifespan. With the best-fit values for the parameters of the model, we find that the lifespan of the universe is 33.3 billion years; that is, the universe will end in about 20 billion years. It is encouraging that the lifespan of the universe can be quantitatively determined. 

Of course, the DES/DESI observation remains to be checked. More and better data are expected in the near future and the aDE model will be rigorously tested. If confirmed, the aDE model parameters (and the universe's lifespan) will be more precisely determined. 

In this paper, we briefly review the aDE model and how the 
observational data determines the parameters of the  model. In particular, we discuss the likeliness that $\Lambda <0$. With the best-fit values of the model, we evolve the universe to the future and estimate the time when $a \to 0$, ending the universe in a big crunch.
 We discuss some issues and then conclude with some remarks.

\section{The aDE Model}

The aDE model starts with the potential for the ultra-light axion field $\phi=f\theta$, 
\begin{equation}
V (\theta) = m_\phi^2f^2\left[1- \cos \left(\theta\right)\right] \;, 
\label{Eq:axion_potential}
\end{equation}
where $f$ is the so called decay constant and $\pi > \theta \ge 0$.  
If the universe starts at a random initial value $\theta=\theta_i \neq 0$, inflation will then lead to $\theta(\bm{x})=\theta_i=\phi_i/f$ everywhere. 
When the Hubble parameter $H \gg m_\phi$, the universe is essentially frozen at the misaligned initial
state $\phi_i$ and $V(\theta_i)$ contribute to the dark energy, {\it i.e.}, the equation of state~(EoS) $w=-1$. As $H \lesssim m_\phi$, $\phi$ starts to roll down along the potential towards $\phi=0$ and deposits the $\phi$ vacuum energy density into matter density (which drops like $a^{-3}$ where $a$ is the scale factor). The current dark energy is a combination of the cosmological constant $\Lambda$ and the axion field. 
The cosmological evolution is governed by Friedmann equations, which are
\begin{align}
\mathcal H^2 & = g(a) + \Omega_\Lambda + \Omega_\phi\;, \nonumber\\
\mathcal H' & = \frac{1}{2} \frac{dg(a)}{d\log a} -\frac{3}{2}  \left(\Omega_\Lambda + \Omega_ \phi \right) \left( w_{\rm DE} +1\right)\;,
\label{eq:Friedmann}
\end{align}
where the dimensionless Hubble parameter is defined as $\mathcal H = H/H_0$, and $H_0$ is the Hubble constant today.
The prime indicates the derivative with respective to the dimensionless cosmic time $H_0 t$ and contributions with constant EoS $w_i\neq-1$ are collectively expressed as 
\begin{equation}
g(a) = \sum_i \frac{\Omega_i}{a^{3(1+w_i)}}\;.
\end{equation}
%
Here the dimensionless energy density is $\Omega_i = \rho_{i,0} /3 H_0^2 M_{\rm Pl}^2$, where $\rho_{i,0}$ is the energy density today at $a=1$; so $\Omega_\Lambda$ is the energy density due to $\Lambda$.
For our purposes here, we consider only $g = \Omega_{\rm m}/a^3$, which includes both the baryon and the dark matter densities.
(It is straightforward to include the radiation or other possible energy content.)
The axion (dimensionless) energy density $\Omega_\phi$ is
\begin{equation}
    \Omega_\phi = \frac{1}{6} \beta^2 \theta'^2 + \overline V(\theta) \;,\quad \overline V(\theta) = \frac{\beta^2 m_\phi^2}{3 H_0^2}(1-\cos\theta)\;,
    \label{eq:Omega_phi}
\end{equation}
where $\beta = f/M_{\rm Pl}$, with $M_{\rm Pl}=2.43 \times 10^{18}$ GeV. Today at $a=1$, we have 
\begin{equation}
 \Omega_{\rm m} +\Omega_{\phi,0} + \Omega_\Lambda = 1\;.
\end{equation}
The axion field evolution is determined by its equation of motion, which is
\begin{equation}
    \theta'' + 3\mathcal H \theta' + \frac{m_\phi^2}{H_0^2}\sin \theta = 0\;.
    \label{eq:EoM_phi}
\end{equation}
Accordingly, the formula for the EoS of the dark energy $(\Omega_\Lambda + \Omega_\phi)$ is 
\begin{equation}
    w_{\rm DE} \equiv \frac{\beta^2 \theta'^2 /6 - \overline V(\theta) - \Omega_\Lambda}{\beta^2 \theta'^2/6 + \overline V(\theta) + \Omega_\Lambda}\;.
    \label{eq:DE_EoS}
\end{equation}
Note that $w_{\rm DE}=-1$ if the axion is absent, or if $H >m_{\phi}$ when the axion field is frozen.
%
As proposed in ref.~\cite{Luu:2025fgw},
it is well approximated by a 2-parameter formula for the $a\leq 1$,
\begin{equation}
    w_{\rm DE}(a) =
    \begin{cases}
    -1 + w_1 (a-a_1)^2 & 1 \ge a \ge a_1   \\
     -1 & a<a_1
    \end{cases} \; ,
    \label{Eq:wfit}
\end{equation}
Note that $w \ge -1$ throughout.
Parameters $a_1$ and $w_1$ can be extracted from data.

\section{Best-fit values for the aDE model}

There are 4 parameters in the model, namely $m_{\phi}$, $f$ or $\beta$, $\Lambda$ or $\Omega_{\Lambda}$, and the initial $\theta_i$ or $\phi_i$. 
The evolution of the axion and $w_{\rm DE}$ requires the knowledge of the (ordinary plus dark) matter density $\Omega_{\rm m}$ and the value of today's Hubble parameter $H_0$.  In the updated version of ref.~\cite{Luu:2025fgw}, with $f=M_{\rm Pl}$, the BAO and SN data from DES~\cite{DES:2025bxy} in combination with others are used to constrain the 5 parameters ($m_\phi, \phi_i, \Omega_m, \Omega_\Lambda, H_0$), while $\Omega_\phi (m_\phi, \phi_i)$~\eqref{eq:Omega_phi} is determined.

\begin{figure}
    \centering
    \includegraphics[scale=0.6]{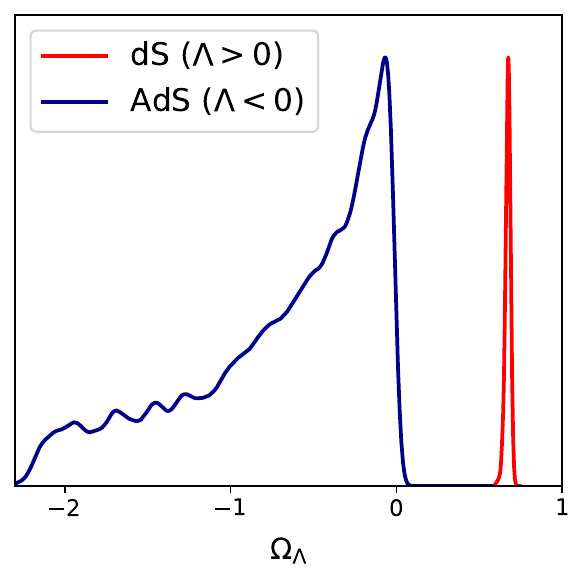} \qquad \includegraphics[scale=0.6]{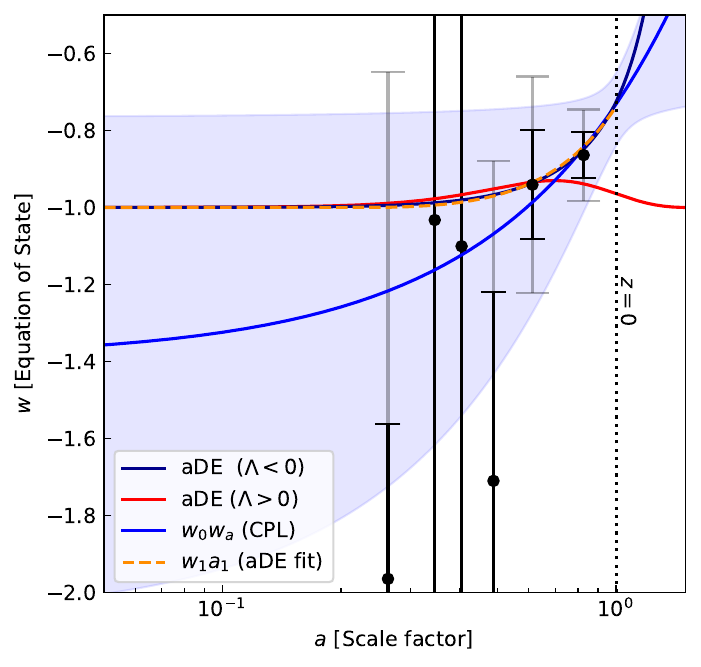}
    \caption{{\it Left:} The marginalized posterior distribution of $\Omega_\Lambda$ obtained by fitting the aDE model with a combination of BAO+SN+BBN+$\theta_*$+$t_U$ data, as described in ref.~\cite{Luu:2025fgw}. The blue curve is for the AdS $\Omega_\Lambda \le 0$ case and the red curve is for the dS $\Omega_\Lambda > 0$ case. The dark blue curve shows that  the value of $\Omega_\Lambda$ is relatively unconstrained with the current DES data. {\it Right:} The EoS in each model. Here we show the best-fit EoS of the $w_0w_a$ model as the blue solid curve for comparison with the aDE ones in dark blue (AdS) and red (dS). The blue shaded region denotes $1\sigma$ uncertainties of this CPL $w_0w_a$ curve. Note that the $w_1-a_1$ curve in dashed orange, as given by eq.~\eqref{Eq:wfit}, overlaps completely with the AdS curve for $a \le 1$. We also show the black data points with $1\sigma$ and $2\sigma$ error bars, which represent the non-parametric constraints of $w$ from DESI+CMB+Union3~\cite{DESI:2025fii} at 6 redshift bins for an easy comparison of how these best-fit curves would fit with DESI data instead of DES.}
    \label{fig:Omega_Lambda} 
\end{figure}

The analysis aims to maximize the likelihood that observational data can be described by a background cosmology driven by the aDE model. Due to the limited available data, there is a relatively large degeneracy in the parameter space. In figure~\ref{fig:Omega_Lambda} (left panel), 
only the posterior distribution of $\Omega_\Lambda$ is shown. It is convenient to separate the $\Omega_\Lambda > 0$ side from the $\Omega_\Lambda \le 0$ side~\footnote{The distribution in the AdS case should be strictly cut off at $\Omega_\Lambda = 0$, but the numerical package used (\texttt{getdist}) (mistakenly) extrapolates the edge to a slightly positive value to ensure normalizability. We have independently checked the MCMC simulations where $\Omega_\Lambda$ can run from AdS to dS. However, because this distribution is multi-modal with a strong attractor at $\Omega_\Lambda \sim 0.7$, these runs will end up getting stuck at the dS solution without exploring the AdS region, even though the latter actually yields a better fit (see Tab. I in ref.~\cite{Luu:2025fgw}). This is the reason we choose to separate these two scenarios.}. For $\Omega_\Lambda > 0$, the best-fit value is consistent with $\Omega_{\Lambda}\lesssim 0.7$, implying that the axion field is not needed and $w = -1$ is acceptable. However, the matter density $\Omega_m \gtrsim 0.3$ seems to be on the high side.

Since the DES/DESI measurement claims $w \ne -1$ at 4.2 $\sigma$ level, let us
focus the aDE model on the $\Omega_\Lambda \le 0$ region. Here, the posterior distribution of $\Omega_{\Lambda}$ has a highly skewed degeneracy. As a result, the average value differs quite a bit from the best-fit value. In particular, as shown in ref.~\cite{Urena-Lopez:2025rad}, setting $\Omega_\Lambda = 0$ (i.e., the pure single axion case) is consistent with the data. However, it is much more likely that $\Omega_{\Lambda}$ is negative~\footnote{A recent analysis ref.~\cite{Gialamas:2025pwv} independently argues that a
negative  $\Omega_{\Lambda}$ is most likely.} (see Tab. I in \cite{Luu:2025fgw} for a comparison of the best-fit values in the aDE model with $\Omega_\Lambda > 0$ and $\Omega_\Lambda < 0$).

It is important to emphasize that the aDE model with an AdS vacuum is more preferred by the recent 
DES data than with a dS vacuum, as demonstrated in figure \ref{fig:Omega_Lambda} with its best-fit EoS, where we also display the CPL/$w_0w_a$CDM model (i.e., $w(a)=w_0+w_a(1-a)$)~\cite{Chevallier:2000qy,Linder:2002et}. This result is obtained because the BAO and SN data probed by 
DES seem to prefer dark energy with EoS deviated from $w = -1$ at late times (from $z \sim 1$ towards $z = 0$). This behavior is better realized when $\Omega_\phi$ becomes the dominant factor compared to $\Omega_\Lambda$ initially, as explained in our previous work~\cite{Luu:2025fgw}. In that study, we also explicitly show that the maximum likelihood (or equivalently $-\chi^2$) of the best-fit AdS model is noticeably higher than that of the dS model. (See ref.~\cite{Luu:2025fgw} for more details about the formal data analysis.) Here we simply quote the best-fit values found for the aDE model in terms of the most relevant cosmological parameters as follows,
\begin{align}
    m_\phi & = 2.93 \times 10^{-33}~{\rm eV}\;, \nonumber \\
     \phi_i &= 6.28 \times 10^{18}~{\rm GeV}\;,\quad  {\rm or} \quad  \theta_i= 0.82 \pi \;, \nonumber \\
    \Omega_{\phi,0} &= 2.33\;, \quad \Omega_\Lambda = -1.61 \quad \rightarrow \quad \Omega_{\rm DE,0} = 0.72 \;, \nonumber\\
     \Omega_{\rm m} &= 0.28 \;,\quad   H_0 = 67.46~{\rm km}/s/{\rm Mpc}= (14.5~{\rm Gyr})^{-1}\;, \nonumber \\
    w_1 &= 0.44, \quad a_1 = 0.23 \;.\label{eq:best-fit}
\end{align}
Here $\Omega_{\phi,0} = \Omega_\phi(a=1)$ and 
$\Omega_{\rm DE,0} =\Omega_{\phi,0} + \Omega_\Lambda = 0.72$. The EoS of dark energy $w_{\rm DE}$ may be obtained using \eqref{Eq:wfit}, where we find that today $w_0 = w_{\rm DE}(a=1) \simeq -0.738$. 
As shown in figure~\ref{fig:Omega_Lambda}, this set of values provides a fit to the 
DES data that is comparable in quality to that of the $w_0w_a$CDM model~\footnote{This statement only applies for the aDE model with DES data as considered in ref.~\cite{Luu:2025fgw}. Other recent studies (e.g.,~\cite{DESI:2025fii, Wolf:2025jed}) show that the ``minimal thawing dark energy'' model similar to our aDE model with $\Lambda = 0$ is less favored than the $w_0w_a$CDM model when being constrained by DESI data.}. For $a<0.23$, the aDE model yields $w=-1$ while the $w_0w_a$CDM model predicts $w \lesssim -1.3$. This big difference will be tested in the near future. We shall use this set \eqref{eq:best-fit} as a benchmark in calculating the lifespan of our universe. The consequences in using the mean values will be discussed later.

\section{Towards the future}

One can predict the future of our universe based on current observations.
Here we investigate the cosmic future based on the best-fit parameters~\eqref{eq:best-fit} of the aDE model.

\subsection{An approximate model}

It is useful to first consider an approximate but analytic version of the aDE model. Here, the picture is clear and the result yields a good estimate of the lifespan.

For small $\theta$ or $\phi$, the potential $V(\theta)$~\eqref{Eq:axion_potential} reduces to $V(\phi)=m_\phi^2 \phi^2/2$.
Multiplying~\eqref{eq:EoM_phi} by $\theta '$ and averaging over a period of the coherent $\theta$ oscillation, $\langle {\theta '}^2 \rangle=3\Omega_{\phi}$,  
we have
$$
\Omega_{\phi}'= -3 \mathcal H \Omega_{\phi} \quad \to \quad \Omega_{\phi} \sim a^{-3}\;. 
$$
So $\Omega_{\phi}$ behaves like matter (although in the actual situation, $\theta$ is not small and will not reach a full cycle).
We propose an approximate model, which describes the axion energy density as
\begin{equation}
    \Omega_\phi(a) \simeq 
    \begin{cases}
        \Omega_{\phi,0}  & a\leq 1\\
        \Omega_{\phi,0} a^{-3} & a>1
    \end{cases}\;,
    \label{eq:omega_phi_approximation}
\end{equation}
where $\Omega_{\phi,0}$ is a constant.
Starting from $a=1$, we consider the effective $\Lambda$CDM model, where ${\bar \Omega}_{\rm m}$ now includes baryon matter, dark matter as well as the axion matter.

\begin{figure}
    \centering
    \includegraphics[width=10cm]{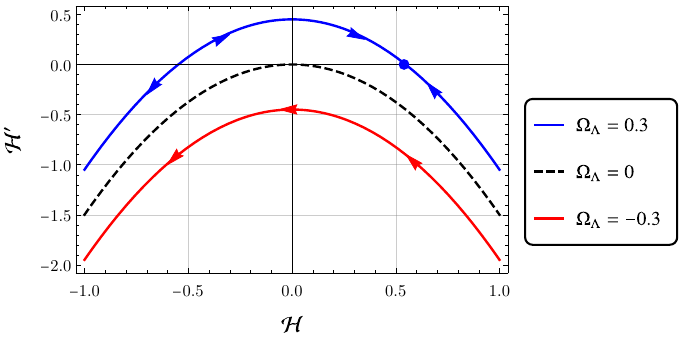}
    \caption{The phase diagram for the Hubble parameter in the effective $\Lambda$CDM model. The arrow indicates the Hubble flow direction.
    The blue dot indicates a stable dS vacuum where the universe will expand exponentially with a constant $\mathcal H = \sqrt \Omega_\Lambda$.
    Here only matter and cosmological constant is included~\eqref{eq:hubble_flow}.}
    \label{fig:hubble_flow}
\end{figure}

The first Friedmann equation is then $\mathcal H^2 = {\bar \Omega}_{\rm m}/a^3 + \Omega_\Lambda$. So 
the consistency relation $\mathcal H^2(a=1) = 1$ indicates that ${\bar \Omega}_{\rm m} + \Omega_\Lambda =1$.
One can combine it with the second Friedmann equation~\eqref{eq:Friedmann} and obtain a phase space relation for the Hubble parameter,
which is~\footnote{It is straightforward to include a radiation or other constant-EoS energy content.}
\begin{equation} 
\mathcal H' = -\frac{3}{2} \left( \mathcal H^2 - \Omega_\Lambda \right)\;.
    \label{eq:hubble_flow}
\end{equation}
The sign of $\Omega_\Lambda$ is crucial for the future. As shown in figure~\ref{fig:hubble_flow}, there are 3 possibilities:

\begin{itemize}
    \item For dS vacuum ($\Omega_\Lambda>0$), one has a stable point where $\mathcal H=\sqrt{\Omega_\Lambda}>0$ and $a \sim e^{Ht}$.
    \item For the Minkowski limit ($\Omega_\Lambda =0$), the universe will approach the point $\mathcal H=0$ with a vanishing $\mathcal H'$. This process will take infinite cosmic time and $a$ continues to grow without bound.
    \item For the AdS ($\Omega_\Lambda<0$) case, there is no such stable configuration. Here, when the Hubble parameter reaches zero (and $a$ reaches its maximum), its derivative $\mathcal H'$ is negative. This means that the Hubble parameter will become negative, where $a$ begins to collapse and starts the crunching phase. It is easy to see that $\mathcal H' <0$ in the actual aDE model with $\Omega_\Lambda<0$.
\end{itemize}

For $\Omega_\Lambda < 0$ in this approximate model, the evolution of the scale factor can be analytically solved from $a'=\mathcal H a$ and eq.~\eqref{eq:hubble_flow}, where the age of the universe $t_0 = 13.8$~Gyr,
\begin{equation}
    a(t)^3 = \frac{\Omega_\Lambda-1}{\Omega_\Lambda} \sin^2\left[\frac{3}{2}\sqrt{-\Omega_\Lambda} H_0 \left(t - t_0 \right) + \arctan(\sqrt{-\Omega_\Lambda}) \right]  \;,
    \label{eq:a}
\end{equation}
Here $a(t=t_0)^3= (|\Omega_\Lambda|+1) \sin^2 \left[ \arctan(\sqrt{|\Omega_\Lambda|}) \right]/|\Omega_\Lambda| =1 $
and we shall take $a(t_0) = 1$ as the starting point. 
The scale factor $a$ reaches a maximum 
\begin{equation}
    a_{\rm max} = \left( \frac{\Omega_\Lambda-1}{\Omega_\Lambda}\right)^{1/3} \;
\end{equation}
and then approaches zero. 
Note that in this approximation, we treat the axion as matter density for $t>t_0$ (before which it is dark energy) till the end~\eqref{eq:omega_phi_approximation}. Therefore, the crunching phase is not simply time reversal of the expanding phase.

The lifespan can be estimated as one cycle of eq.~\eqref{eq:a}, when $a \to 0$,
\begin{equation}
  T (\Omega_\Lambda<0) = \frac{1}{H_0}\left( \frac{2\pi-2 \arctan{\sqrt{|\Omega_\Lambda|}}}{3 \sqrt{|\Omega_\Lambda|}} \right) + t_0 \;.
    \label{eq:T_analytical}
\end{equation}
By taking the best-fit data $\Omega_\Lambda = -1.61$~\eqref{eq:best-fit}, one obtains $a_{\rm max}=1.17$ and a universe of lifespan $T \simeq 30.9$~Gyr.
The crunching phase takes up $\sim 39\%$ of the total lifespan.

\subsection{Estimating the lifespan}

As indicated in the best-fit parameters~\eqref{eq:best-fit}, $\Omega_\Lambda = -1.61$ and the mass of the axion is roughly $m_\phi \simeq 2.04 H_0$. So the axion is still rolling towards its minimum in the first period today ($a=1$).
With the initial angle at $\theta_i= 0.82 \pi$, the evolution of $\Omega_\phi$ cannot be simply approximated as matter-like. One has to numerically solve the dynamics.
Taking the $a(H_0 t_0 = 0.952)=1$ as the starting point, we fix the initial conditions from the consistency of the Firedmann equation $\mathcal H^2(a=1)=1$ and the implication from current observation on today's EoS $w_0 = w_{\rm DE}(a=1)=-0.74$.
One obtains, from eq.~\eqref{eq:Friedmann} and eq.~\eqref{eq:DE_EoS},
\begin{align}
\overline V(\theta_0) & = \frac{1}{2}(1- \Omega_{\rm m})\left(1- w_0\right) - \Omega_\Lambda \;,\nonumber \\
\theta_0' &= \pm \sqrt{ \frac{3}{\beta^2} (1-\Omega_{\rm m})(1+w_0)} \;,
\label{eq:initial_conditions}
\end{align}
where $\theta_0 = \theta(a=1)$.
Here, the sign of the derivative depends on the position $\theta_0$.
Since the axion in our scenario is still rolling down towards its minimum for the first time, we take $\pi> \theta_i > \theta_0>0$ and $\theta_0'<0$ in the following.

\begin{figure}
    \centering
    \includegraphics[width=7cm]{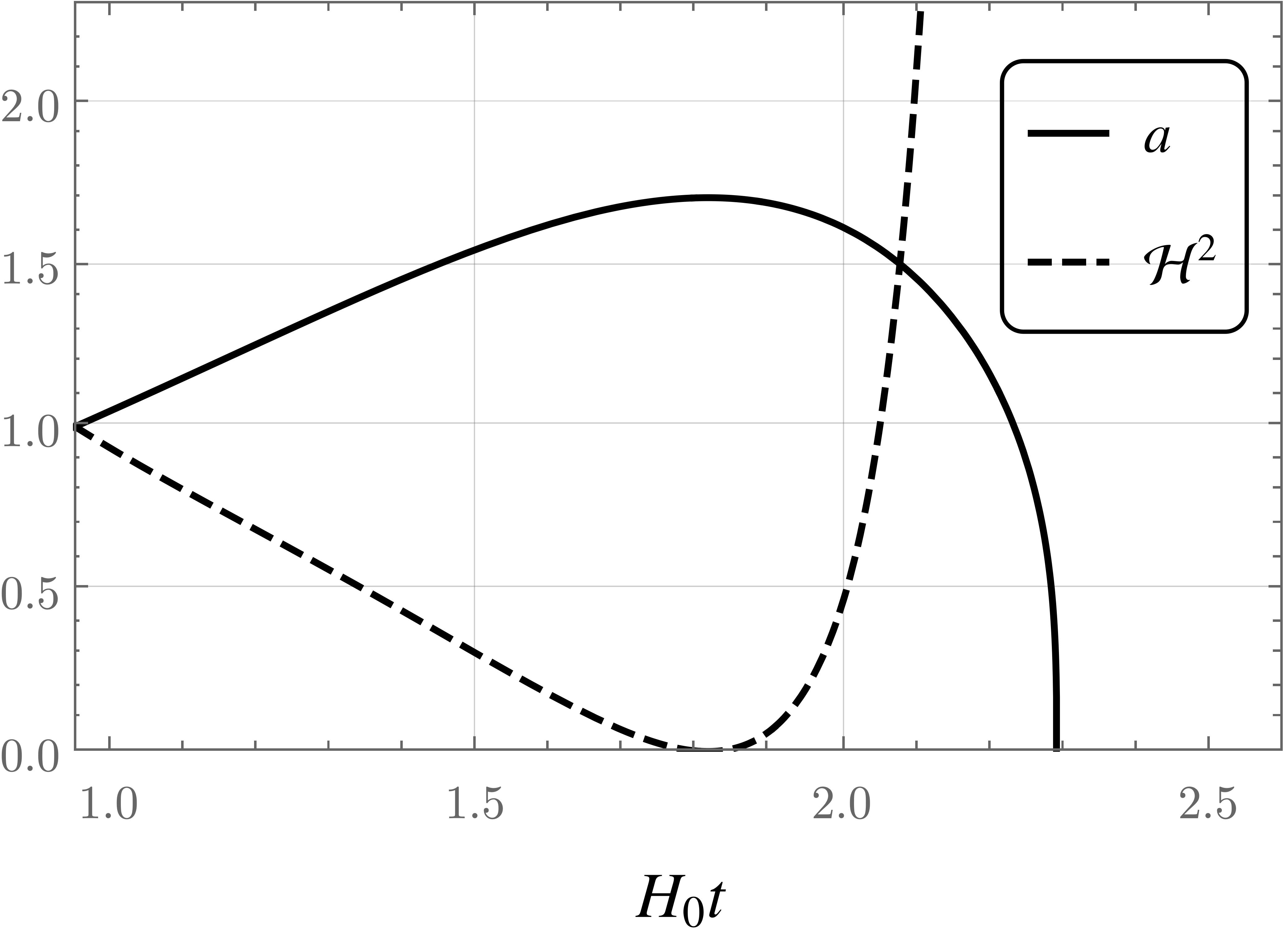}
    \includegraphics[width=7cm]{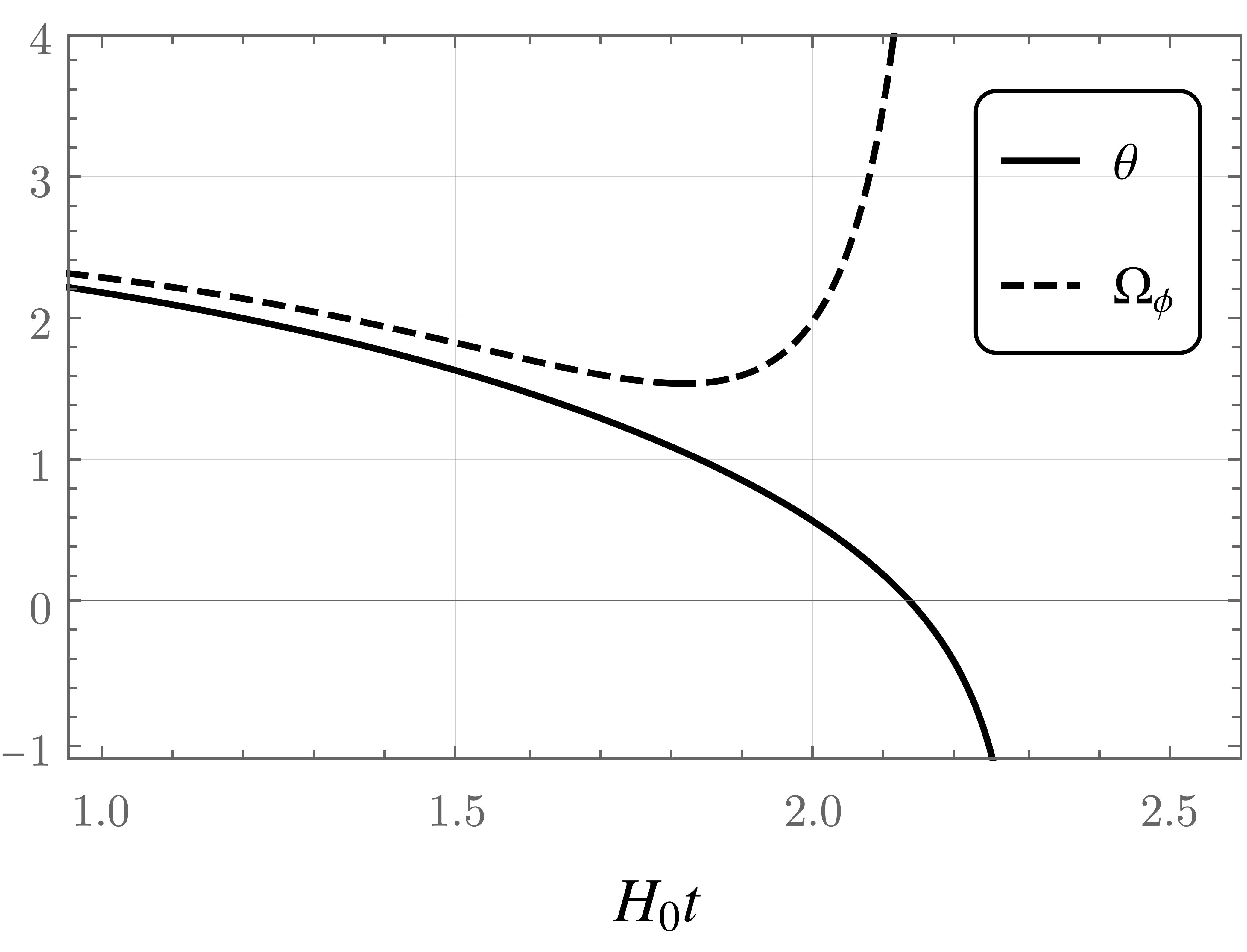}
    \caption{{\it Left Panel}: The cosmic evolution of the scale factor and the Hubble parameter. {\it Right panel}: The cosmic evolution of the axion field value and its (dimensionless) energy density. They are all calculated with the best-fit parameters~\eqref{eq:best-fit}. The starting point is $H_0 t_0 = 0.952$.}
    \label{fig:best-fit_evolution}
\end{figure}

\begin{figure}
    \centering
    \includegraphics[width=9cm]{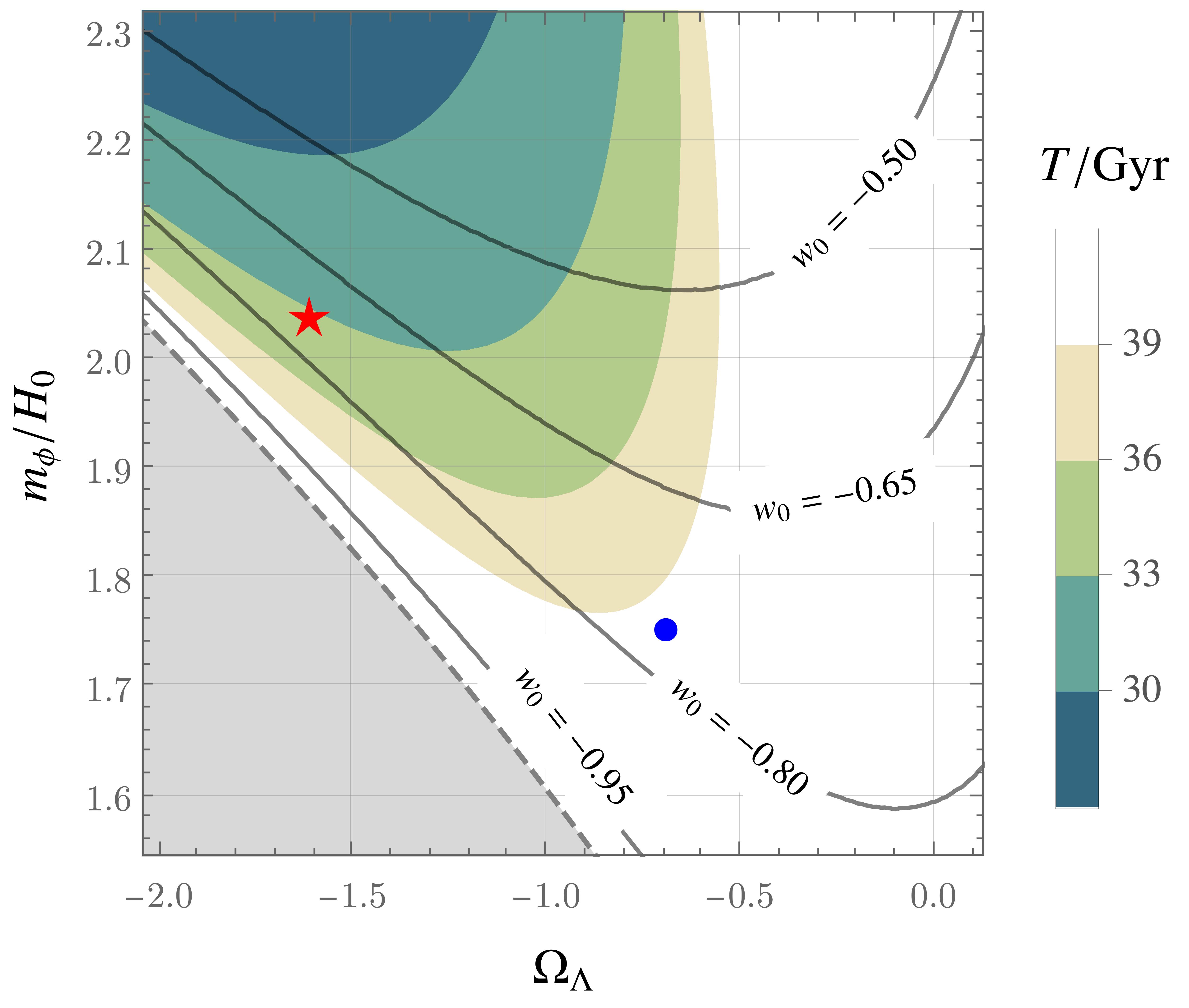}
    \caption{The lifespan $T$ predicted by aDE model in the $m_\phi$-$\Omega_\Lambda$ plane with fixed $f=M_{\rm Pl}$ and $\Omega_{\rm m} = 0.284$. Here the black contours are the associated dark energy EoS $w_0$ today. The gray-shaded region is not allowed by the aDE model.
    The red star labels the best-fit parameter from current observation~\eqref{eq:best-fit} and the blue dot labels the mean value~\eqref{meanT}.}
    \label{fig:m_Lambda}
\end{figure}

The initial conditions determine the evolution of the axion via eq.~\eqref{eq:EoM_phi}.
Meanwhile, the evolution of the scale factor is governed by
\begin{align}
    a' & = \mathcal H a \;, \label{eq:EoM_a}  \\
    \mathcal H' & = -\frac{3}{2}\frac{\Omega_{\rm m}}{a^3} - \frac{1}{2} \beta^2 \theta'^2 \nonumber \;.
\end{align}
This coupled system eq.~\eqref{eq:EoM_phi} and eq.~\eqref{eq:EoM_a} can be solved numerically. Technically, to avoid the singularity, we set the threshold of the minimum scale factor as $a_{\rm min} = 10^{-3}$, which terminates the calculation when $a<a_{\rm min}$ and set the lifespan $T \simeq t_{\rm max}$, where $a(t_{\rm max}) \simeq a_{\rm min}$.~\footnote{Choosing a specific $a_{\rm min}$ is for the numerical purpose. Setting for example $a_{\rm min}=10^{-2}$ leads to numerical termination at the same $H_0t \simeq 2.29$, which makes no difference up to two decimal points. This is due to the rapid drop of $a$ towards the end, as shown in figure~\ref{fig:m_Lambda}.}
Taking the best-fit parameter~\eqref{eq:best-fit}, we obtain the cosmic evolution as shown in figure~\ref{fig:best-fit_evolution}.
The scale factor reaches the maximum $a_{\rm max} \simeq 1.69$ at $H_0 t \simeq 1.71$. Meanwhile, the Hubble parameter hits the zero at the same time and turn negative afterwards, which causes the crunch. 
This happens roughly $11$ billion years from now.

The numerical evaluation terminates at $H_0 t \simeq 2.29$ and this gives us a lifespan of our universe, 
\begin{equation}
T\simeq 33.3 ~{\rm Gyr}\;.
\label{T}
\end{equation}
One can see that the crunching phase only takes up $\sim 25\%$ of the total lifespan.
Comparing to the analytical approximation~\eqref{eq:T_analytical}, the onset oscillation delays the moment of reaching the $a_{\rm max}$, and the kinetic domination of the axion accelerates the crunching phase. The net effect gives a slightly larger lifespan.

We also scan the parameter space and obtain the corresponding lifespan as shown in figure~\ref{fig:m_Lambda}. 
The consistency relation $\mathcal H^2(a=1)=1$ can reduce one degree of freedom,
which fixes $\theta_i$ or $\phi_i$ by other parameters.
Therefore, the dark energy EoS today is determined $w_0 (\Omega_{\rm m},\Omega_\Lambda,m_\phi,\beta)$, which are shown as black contours in figure~\ref{fig:m_Lambda}. This in turn gives the initial conditions~\eqref{eq:initial_conditions} that is used to solve for the future. 
The lifespans are shown as colored bands.
Note that the analytical estimation~\eqref{eq:T_analytical} only depends on the $\Omega_\Lambda$, which would manifest itself as vertical contours in the figure~\ref{fig:m_Lambda} and deviate from numerical evaluation in the small mass region.


\section{Discussion}

We give some discussion in the following.

\begin{itemize}
    \item Note that the DESI (northern hemisphere) measurements are in good agreement with the DES (southern hemisphere) data set, so combining them yields the $w \ne -1$ result at $4.2 \sigma$ significant level. As shown in figure \ref{fig:Omega_Lambda}, the best-fit dark blue curve (\ref{eq:best-fit}) for the DES data set also fits well the DESI data set.
    \item To gain a qualitative understanding on why a negative $\Omega_\Lambda$ is preferred, let us look at the redshift bin from DESI with the smallest error:  $w \simeq -0.86$ at $a \simeq 0.77$ as shown in figure \ref{fig:Omega_Lambda}. The best fit red curve (i.e., with $\Omega_\Lambda >0$) is $1 \sigma$ off from this data point. Here, $\Omega_{\phi,0}  < 0.72$  is limited.  Qualitatively, a larger $\Omega_{\phi,0}$ tends to move $w$ closer to the central bin value. That is,  a better fit (to the 2 lowest $z$ bins) requires a larger $\Omega_{\phi,0}$, which leads to a negative $\Omega_\Lambda$ and a big crunch.

    \item As shown in Table 1 and figure 4 in the extended version of ref.~\cite{Luu:2025fgw}, because of the large degeneracy and the highly skewed probability distribution, the best-fit value $\Omega_{\Lambda}=-1.61$ is way outside the $1 \sigma$ range of the mean value
    \begin{align}
        \langle \Omega_\Lambda \rangle =-0.69^{+0.70}_{-0.20}\;.
    \end{align}
    This also shows that, although not favored, $\Omega_{\Lambda}=0$ is consistent with the DESI data~\cite{Urena-Lopez:2025rad}. With the corresponding average mass $\langle m_\phi \rangle=2.51 \times 10^{-33}$ eV and $\langle \phi_i \rangle = 5.22 \times 10^{18}$ GeV, the lifespan will be 
    \begin{align}
        \langle T \rangle = 40.3\, {\rm Gyr} \quad {\rm at} \quad \langle w_0 \rangle  =-0.763
        \label{meanT}
    \end{align}        
    which is longer than $T$ in \eqref{T}. This result is shown as a blue dot in figure~\ref{fig:m_Lambda}.

    \item Since both a radiation and a positive curvature term are negligible today, including them will make little difference to the analysis. However, including a negative curvature term will shorten the lifespan. In fact, a negative curvature term will render the lifespan finite even if the cosmological constant term is absent.

    \item For smaller decay constant $f \lesssim M_{\rm Pl}$, which corresponds to $\beta<1$, it only affects the axion evolution through its effect on the Hubble flow. As shown in the equation of motion~\eqref{eq:EoM_phi}, there is no explicit $\beta$ dependence. However, it enters the Hubble flow via $\Omega_\phi(\beta) = \beta^2 \Omega_\phi(\beta=1)$. Since the axion is still in its early rolling period, the energy density is still dominated by the potential energy, where $\beta$ enters in the combination $m^2_\phi\beta^2$~\eqref{eq:Omega_phi}, reflecting a degeneracy in data fitting. In any case, because of the degeneracy already present, a small lowering of the value of $f$ will make little difference to the fit.

    \item Note that throughout the evolution in the aDE model, $H^2 \ge 0$ (on-shell). If the universe starts in an AdS space, $ H^2 = -|\Lambda| <0$, or more generally starts with $H^2 <0$ (off-shell), which may be approached via a tunneling process \cite{Coleman:1980aw}, its evolution will be very different.

    \item The Hubble tension is lurking at the background of our analysis as well as the DES/DESI analaysis~\cite{Pang:2025lvh}. If we bring in the axi-Higgs model~\cite{Fung:2021wbz,Fung:2021fcj}, $H_0$ can be uplifted by $2\%$--$3\%$, relieving some of the Hubble tension. 
    Since the BAO and SN data are late time measurements, it may be more appropriate to use the late time Hubble value, i.e., $h=0.71$ instead of $h=0.67$. That is, $H_0$ shifts from $(14.5\, {\rm Gyr})^{-1}$ to $(13.8\, {\rm Gyr})^{-1}$, which will shorten the lifespan by about 1 billion years.

    \item If there is another ultra-light axion with mass $m \le 10^{-34}$ eV, its energy is still contributing to the dark energy density today. As a result, its presence will further lower the value of $\Lambda$. That is, the big crunch will still happen, though the lifespan of our universe will depend on the properties of this additional axion.
        
\end{itemize}

\section{Remarks}

Here, a few remarks on the overall picture are in order. \\

{\bf String Theory Implications} --- String theory suggests that constructing a meta-stable vacuum with a positive cosmological constant can be challenging, while constructing vacua with a negative cosmological constant is generic~\cite{Danielsson:2018ztv,Obied:2018sgi,Ooguri:2018wrx}. Presumably, non-supersymmetric vacua are unstable or meta-stable, so they will eventually evolve towards supersymmetric vacua, which have negative $\Lambda$. 

It is clear that SUSY is broken today, as not a single super-partner has been discovered, suggesting that the super-partners have masses of order of TeV.
A generic mass splitting term of TeV scale will shift the standard model particle masses, but cosmological observations such as the CMB measurements shows that the masses of the electron and nucleons should not have been shifted by more than a few percent since big bang nucleosynthesis time. This suggests, in addition to the standard model particle masses in the supersymmetric vacuum,  there is an additional SUSY-breaking mass term of the form,
$$
V(\phi) (1-p_R)\;,  \quad p_R=(-1)^{3(B - L)+2s}
$$
where the super-partners have odd R-parity, $p_R=-1$, while the standard model particles have $p_R=1$.
This raises only the masses of the super-partners until $V(\phi)=0$ at $\phi=0$, when the super-partner spectrum matches that of the standard model particle spectrum. 
Irrespective on whether the AdS ground state is supersymmetric or not, the universe will pass it at roughly $H_0 t \simeq 2.1$ (see the right panel of Fig.~\ref{fig:best-fit_evolution}) before reaching the crunch.\\

{\bf Small $|\Lambda|$} --- Earlier the grand puzzle is the smallness of $\Lambda$ in de Sitter space. Now, the puzzle is to explain the smallness of the negative $\Lambda$ in AdS space.
Note that the explanation for these 2 cases can be very different. For example, the statistical explanation for an exponentially small positive $\Lambda$~\cite{Sumitomo:2013vla,Qiu:2020los} does not work when $\Lambda <0$. Interestingly, in a set of Calabi-Yau orientifold models in Type IIB string theory, one finds that $\Lambda$ can be exponentially small, only if it is negative in a supersymetric vacuum~\cite{Demirtas:2021ote,Demirtas:2021nlu}. 
This provides a plausible explanation why the negative $\Lambda$ is so small. However, with broken SUSY today, normal quantum corrections to $\Lambda$ that are absent in a SUSY vacuum are in general present now.\\

{\bf Axiverse } --- As noted before, ultralight axions (or axion-like particles) are ubiquitous in string theory~\cite{Svrcek:2006yi,Arvanitaki:2009fg}. They pick up their exponentially small masses via non-perturbative dynamics (cf.~ref.~\cite{Hui:2016ltb}). There is tremendous freedom to introduce any number of ultra-light axions with a spectrum of masses and couplings (i.e., decay constants). Of course, it is much more interesting when the introduction of a specific ultra-light axion resolves some outstanding puzzles in astrophysics and/or cosmology. 

The best known case is the fuzzy dark matter (FDM) model, where an axion of mass around $10^{-22}~{\rm eV}$ is the source of the dark matter~\cite{Hu:2000ke, Schive:2014dra}. To resolve some issues with the diversity of ultra faint dwarf galaxies, the introduction of a second axion of mass around $10^{-20}$~eV seems to be necessary~\cite{Luu:2018afg}. Another even lighter axion of mass around $10^{-29}$~eV introduced in the so-called axi-Higgs model~\cite{Fung:2021wbz,Fung:2021fcj} helps to explain the $^7$Li puzzle in BBN, the Hubble tension \footnote{See also ref.~\cite{Poulin:2018cxd} for another possibility.} and the isotropic cosmic birefringence~\cite{Minami:2020odp,Fujita:2020ecn,Nakagawa:2025ejs}. The aDE model introduces one with mass $10^{-33}$~eV. 
These 4 ultra-light axions have comparable decay constant $f \lesssim M_{\rm Pl}$. Presumably, they are Ramond-Ramond modes in string theory under the compactification to 3 spatial dimensions. If their existences are confirmed, these axions together with their properties will shed valuable insight into the string theory compactification.\\

{\bf Astrophysical/Cosmological Implications} --- In view of the dynamical dark energy impact on the evolution of the universe, it is interesting to re-analyze some of the dynamics involving astrophysical phenomena. As $a \to 0$, matter is being squeezed together. This is expected to enhance the formation of black holes, in particular the merging of black holes. Eventually, it is plausible that the universe ends in giant black holes. Including a growing radiation density as $a \to 0$ will make little difference.

Note that the scale factor $a$ has a periodic behavior and we consider only one cycle for our universe. As mentioned earlier, because of the nature of the axion field, the process is irreversible. Although very unlikely, one cannot rule out that, in the presence of quantum effects, there is a way for the universe to transition to the next cycle. However, this reincarnation will lead to a very different universe from our present one, not to a cyclic universe. 

There are 3 types of singularities we encounter in the study of gravity : black holes, the origin of the universe and the crunch at the end of the universe. As the universe is collapsing, one can imagine that matters are push together to form a giant black hole, which in turn shields/hides the crunch singularity.
 
On the local level, the possible collision of our Milky Way with its largest neighbor, the Andromeda galaxy, is predicted to occur in about 4 to 10 billion years~\cite{cowen2012andromeda,sawala2025no}. It is interesting to re-estimate the time of their collision (or no collision) before the big crunch.

\section{Summary}

We study a very simple axion plus $\Lambda$ model with
$\Omega_{\rm m} + \Omega_{\phi,0} + \Omega_{\Lambda} =1$,
where the matter density $\Omega_{\rm m} =0.28$ includes both the ordinary matter and the dark matter densities, while the dark energy density includes the axion energy density and the cosmological constant: $\Omega_{\phi,0} + \Omega_{\Lambda}
=0.72$. Here, introducing a radiation density makes little difference. For the dark energy EoS $w= -1$, $\Omega_{\phi}=0$. For $w> -1$ and varying, the ultralight axion in the aDE model can explain its varying behavior, provided that $w \ge -1$. In fitting the DES data in combination with other constraints from CMB, BBN and age of the universe, we determine the axion mass to be $m_\phi\simeq 2.93 \times 10^{-33}$~eV and $\Omega_{\phi,0} \simeq 2.33$, so $\Omega_{\Lambda}\simeq -1.61$. This negative cosmological constant means that our universe contains the matter and the $\phi$ energy in an AdS space, resulting in its eventual contraction, ending in a big crunch. Defining the big crunch as the end of our universe allows us to estimate the lifespan of our universe to be about 33.3 billion years. 

We emphasize that the determination of the lifespan of our universe depends on the recent observation that $w > -1$ at small redshift and the validity of the aDE model. It is crucial that the DES/DESI observation is confirmed and the aDE model is rigorously tested. Fortunately, a number of projects measuring different aspects of the dark energy are forthcoming in the near future. We look forward to a more precise determination of the universe's lifespan.

\section*{Acknowledgment}
We thank Dennis Overbye and Gary Shiu for discussions.
The work of Y.~-C.~Qiu is supported by the K.~C.~Wong Educational Foundation.

\bibliographystyle{JHEP} 
\bibliography{reference}

\end{document}